\documentclass[conference]{IEEEtran}
\IEEEoverridecommandlockouts
\usepackage{cite}
\ifCLASSINFOpdf
   \usepackage[pdftex]{graphicx}
\else
\fi
\usepackage{url}
\usepackage{amsmath,amssymb,amsfonts}
\usepackage{algorithmic}
\usepackage{graphicx}
\usepackage{hyperref}
\usepackage{textcomp}
\usepackage{xcolor}
\usepackage[separate-uncertainty=true,multi-part-units=single]{siunitx}
\def\BibTeX{{\rm B\kern-.05em{\sc i\kern-.025em b}\kern-.08em
    T\kern-.1667em\lower.7ex\hbox{E}\kern-.125emX}}
    
\usepackage{scalerel}
\usepackage{tikz}
\usetikzlibrary{svg.path}

\ifdefined\qtyproduct
\else
  \ifdefined\NewCommandCopy
    \NewCommandCopy\qtyproduct\SI
  \else
    \NewDocumentCommand\qtyproduct{O{}mm}{\SI[#1]{#2}{#3}}
  \fi
\fi

\definecolor{orcidlogocol}{HTML}{A6CE39}
\tikzset{
  orcidlogo/.pic={
    \fill[orcidlogocol] svg{M256,128c0,70.7-57.3,128-128,128C57.3,256,0,198.7,0,128C0,57.3,57.3,0,128,0C198.7,0,256,57.3,256,128z};
    \fill[white] svg{M86.3,186.2H70.9V79.1h15.4v48.4V186.2z}
                 svg{M108.9,79.1h41.6c39.6,0,57,28.3,57,53.6c0,27.5-21.5,53.6-56.8,53.6h-41.8V79.1z M124.3,172.4h24.5c34.9,0,42.9-26.5,42.9-39.7c0-21.5-13.7-39.7-43.7-39.7h-23.7V172.4z}
                 svg{M88.7,56.8c0,5.5-4.5,10.1-10.1,10.1c-5.6,0-10.1-4.6-10.1-10.1c0-5.6,4.5-10.1,10.1-10.1C84.2,46.7,88.7,51.3,88.7,56.8z};
  }
}
\newcommand\orcidicon[1]{\href{https://orcid.org/#1}{\mbox{\scalerel*{
\begin{tikzpicture}[yscale=-1,transform shape]
\pic{orcidlogo};
\end{tikzpicture}
}{|}}}}
    
\usepackage[bottom=0.75in,top=0.68in, right=0.7in, left=0.7in]{geometry}
\usepackage{subcaption}

\begin{document}

\title{Comprehensive Design Validation of \\ serially powered CMS Phase-2 Pixel Modules

\thanks{ \textbf{Acknowledgment}
The work presents the author's personal results which have been incorporated into the Quality Control for the CMS Inner Tracker.}
}
\author{G. Bonomelli$^{1}$ \orcidicon{0009-0003-0647-5103} on behalf of the CMS Tracker Group
\vspace{0.1cm}
\\
\small{
$^{1}$ETH, Zürich, Switzerland
}
}

\maketitle

\begin{abstract}
After the Large Hadron Collider (LHC) upgrade into High Luminosity LHC (HL-LHC), the instantaneous luminosity is expected to reach values up to $7.5 \times 10^{34} cm^{2}/s$, causing a harsher radiation environment as well as a significant increase in data rate. The current CMS Tracker detector would not be able to operate under these conditions and it will be replaced by an upgraded version known as Phase-2 \cite{cmsPhase2}. 
In view of the detector upgrade and as part of the design validation process, a Quality Control (QC) test flow has been developed to characterize the first pixel modules prototypes and evaluate their performance. The results of this procedure were the starting point for small design adjustments, especially for the the High Density Interconnect or HDI, the flexible low mass PCB that distributes power and signals to the module and controls the readout through a high speed data transmission channel.
\\
This talk includes qualification tests performed on the CMS Phase-2 design to ensure that all the pixel module components satisfy the upgrade specifications, for example in terms of power consumption and leakage current stability. Additionally, stress tests were conducted to probe the limits of the design, demonstrating the robustness and endurance of the module layout. Due to the differing material properties of the HDI copper layers and the silicon sensor and readout chip, temperature gradients induce different thermal expansion and contraction, resulting in mechanical stress on the bump-bond interface. For this reason, among the destructive measurements, dedicated thermal stress tests were carried out to evaluate the bump-bond strength and durability for different bump bonding techniques. 
\end{abstract}
\vspace{0.1cm}

\vspace{-0.15cm}
\section{Introduction}
In the next years, the upgrade of the CERN accelerator complex into High Luminosity LHC will bring major changes. The new system will reach an expected peak luminosity of $7.5 \times 10^{34} cm^{2}/s$, leading to a significantly higher number of interactions per bunch crossing, from the current pile-up value of $\sim 40$ up to 200. As a consequence, the radiation environment will also be significantly harsher than it is at present, with doses up to 1.9 Grad. Since the current CMS Inner Tracker (IT) would not be able to operate under these conditions while keeping the same performance, it will be replaced by the CMS Pixel Phase-2 upgrade before the HL-LHC operation starts, during the Long Shutdown-3. 

\vspace{-0.15cm}
\section{Pixel Module Design}
\begin{figure}
    \centering
    \includegraphics[width=0.28\textwidth]{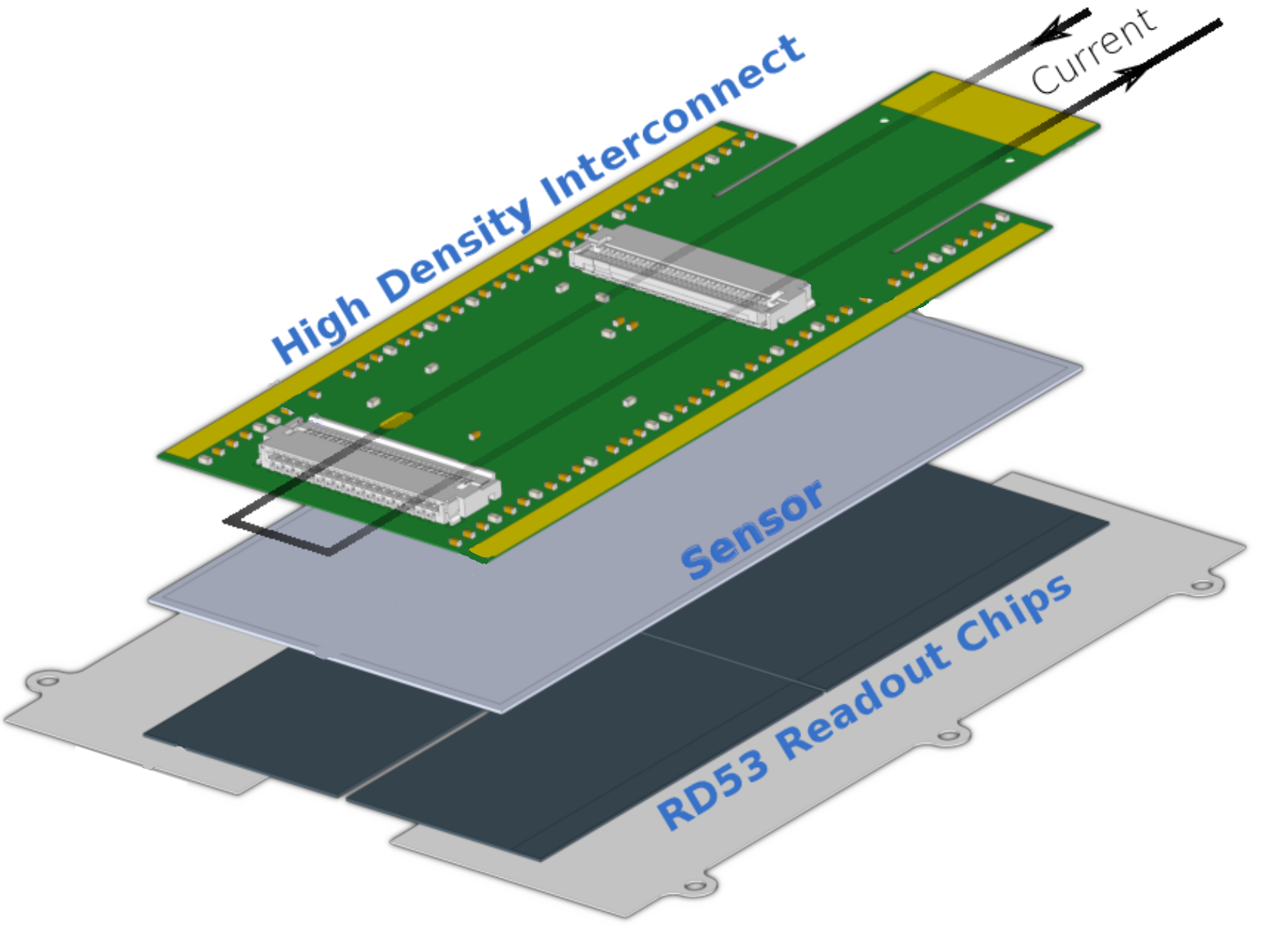}
    \caption{Schematics of the barrel Pixel module prototype for the Phase-2 upgrade.}
    \label{fig:pixel_module}
    \vspace{-0.6cm}
\end{figure}

Stringent requirements are imposed for the IT upgrade, such as radiation tolerance, a large memory buffer and a high readout speed to cope with the very high pile-up values. All those prerequisites are satisfied by a hybrid pixel detector consisting of a readout-chip (ROC), realized in 65 nm CMOS technology, bump bonded to a sensor featuring a pixel size of $25\times100\mu m^{2}$. The silicon sensor is glued to a flexible low mass PCB called High Density Interconnect (HDI), which distributes signals and power to the chips and controls the read-out of the ROCs (Fig. \ref{fig:pixel_module}).

The HDI allows the readout-chips within a module to send data either independently or through different "master-slave" configurations. In the first case, each ROC transmits data using per-chip high-speed channels. The second configuration allows one or more chips to serialize differential data inputs from other chips along with their own data with an internal speed of 320 Mbps. The inputs are combined into a single high speed transmission up to 1.28 Gbps, reducing the amount of cables needed \cite{dinardo}.
In addition, to stay within the specification for the power consumption, the Phase-2 upgrade will foresee a serial powering scheme, a novelty for a large-scale detector. This consists of pixel module units, either dual or quad-chip modules, organized in ladders and sharing the same input current. Compared to the current parallel powering, it will drastically reduce the power consumption of the Inner Tracker as well as the material budget \cite{Malte} \cite{Bane}. 
\\
Within a single module, to prevent one failing chip from breaking the whole chain, the ROCs are powered in parallel through a Shunt Low Drop Out (Shunt-LDO) regulator which consists of two parts: a voltage regulator, providing the necessary supply voltages and in parallel a shunt circuit, which consumes the additional current not used by the LDO \cite{sldo}.

\vspace{-0.15cm}
\section{Design Validation Studies \& Results}
\begin{figure}
    \centering
    \includegraphics[width=0.45\textwidth]{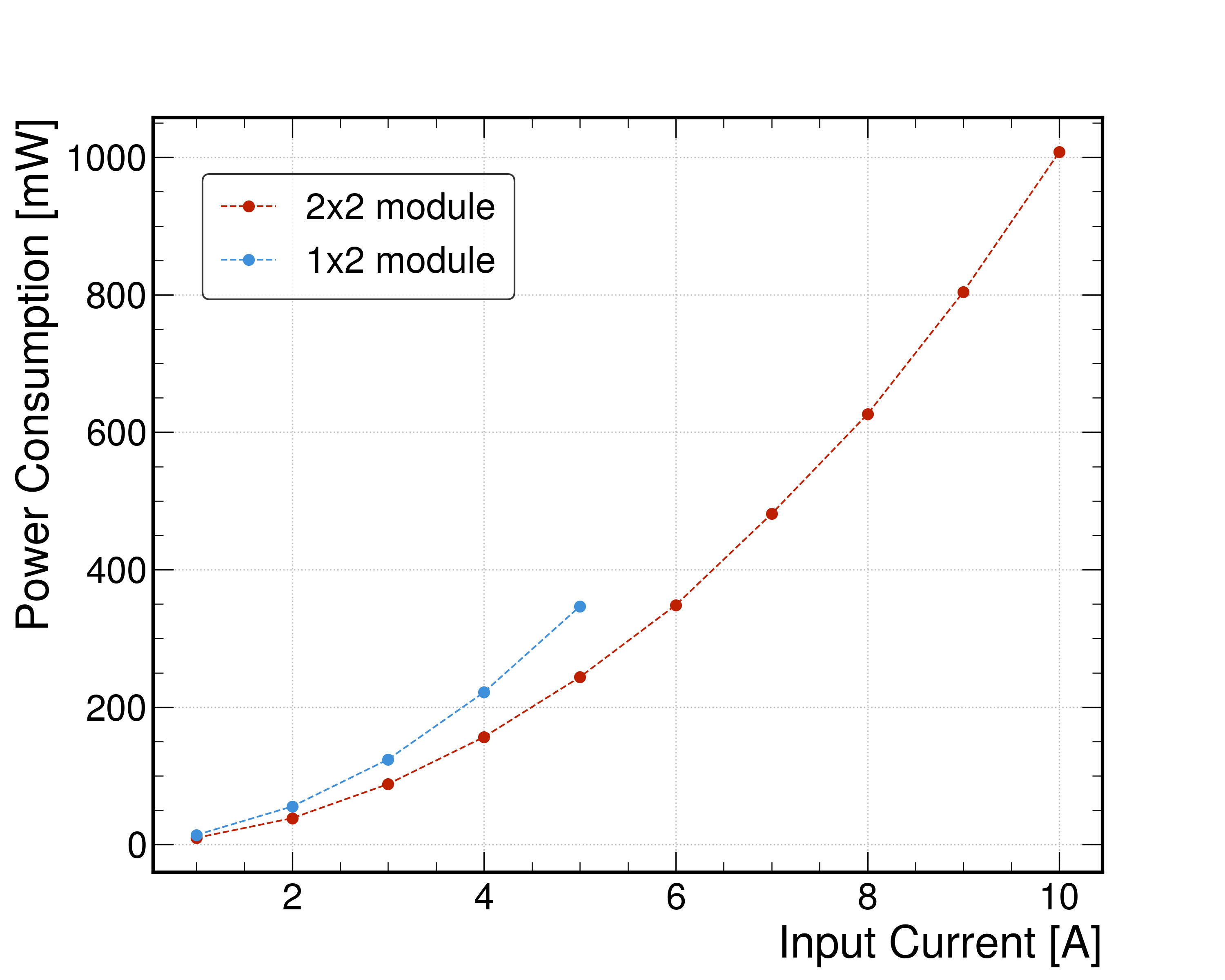}
    \caption{Power consumption for quad (2x2) and dual (1x2) chip modules.}
    \label{fig:power}
    \vspace{-0.6cm}
\end{figure}

\begin{figure}
    \centering
    \includegraphics[width=0.4\textwidth]{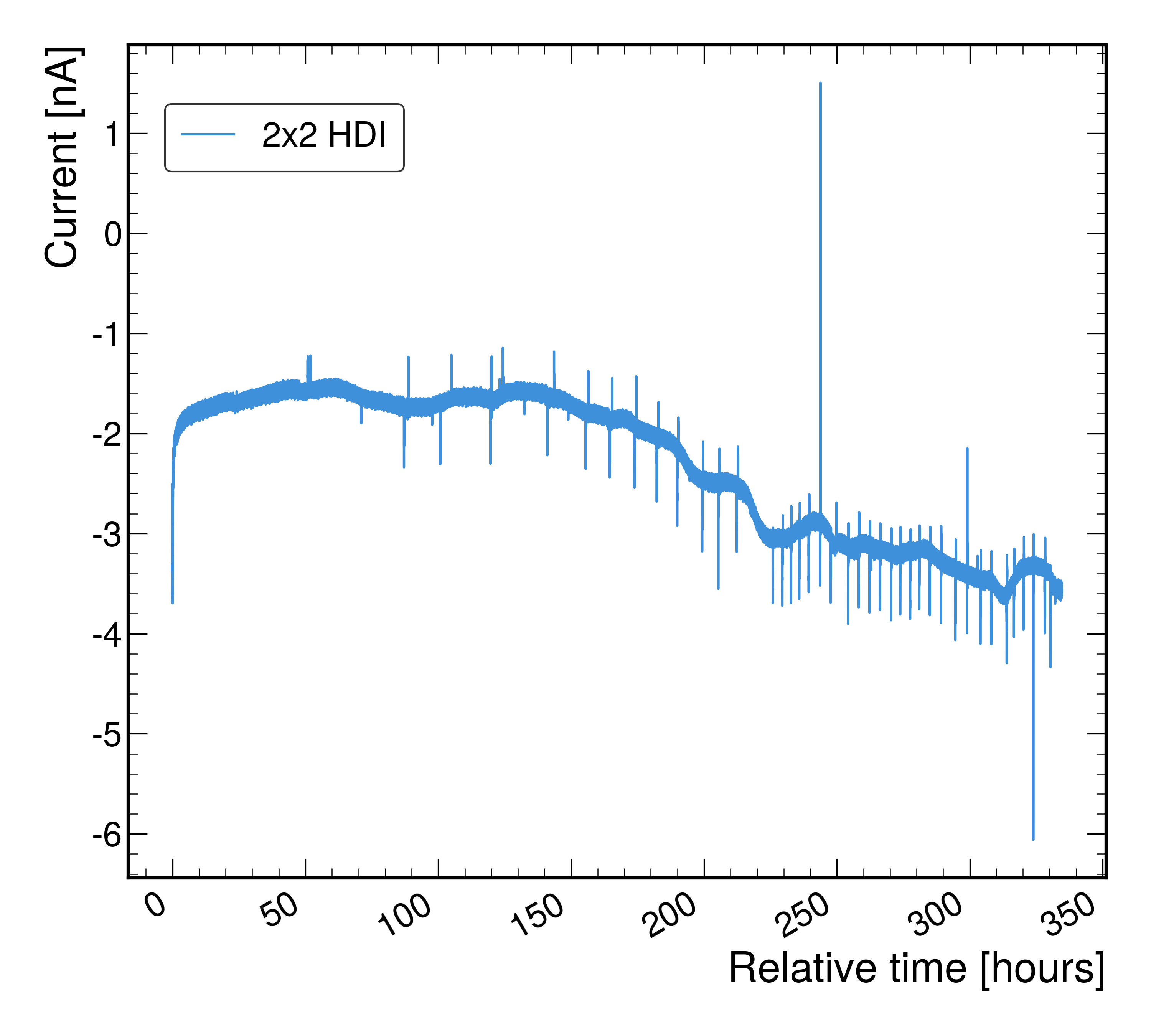}
    \caption{Leakage current stability over time measured on the quad (2x2) HDI design.}
    \label{fig:stability}
    \vspace{-0.6cm}
\end{figure}

\begin{figure}
    \centering
    \includegraphics[width=0.5\textwidth]{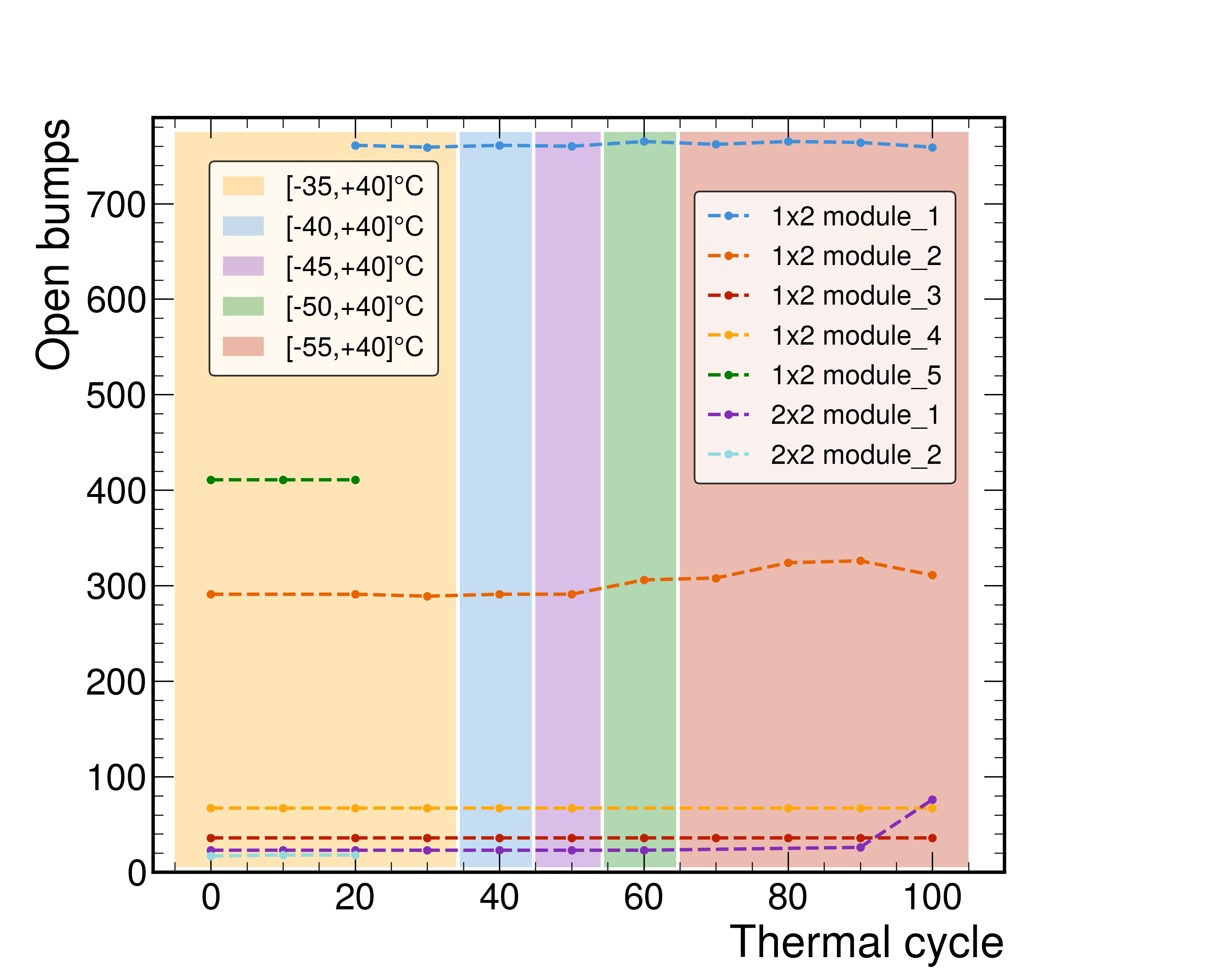}
    \caption{Open bump analysis after thermal stress tests with various extended temperature ranges on dual (1x2) and quad (2x2) prototype modules. }
    \label{fig:thermal_cycles}
    \vspace{-0.6cm}
\end{figure}

To qualify the readout-chip and the HDI designs, several prototype modules underwent the Quality Control (QC), a procedure that has been developed and optimized to evaluate the pixel modules performance. 
Before the assembly, each HDI version was individually tested to identify potential faults in the PCB layout. Specifically, the power distribution network was evaluated by measuring the current draw and the resistance mismatch of the input and output currents on the HDI for different values of supply currents (Fig. \ref{fig:power}). These results are in agreement with simulations performed under the same experimental conditions. Additionally, extended high-voltage tests were conducted to assess leakage current stability over time, with the backside maintained at ground potential, to ensure long-term reliability under operating conditions (Fig \ref{fig:stability}).
\\
On full modules, particular focus was placed on SLDO measurements, and on the power consumption under varying input currents to ensure their agreement with the upgrade standards. Moreover, the data link integrity and transmission capability were tested with all the possible "master-slave" configurations, confirming that the HDI design reliably supports them. 
\\
The results of the first prototypes were crucial in guiding the improvements and modifications to the pixel module design. Once the upgraded and finalized versions were available, the same tests were performed to confirm that the adjusted module design continued to meet the selection criteria. 
Further failure scenarios tests were performed to probe the limits of the module design. HDIs were proven to be able to sustain more than double the default input current, while simulating a short in one or multiple readout-chips. No visible signs of burnt marks or operation problems in the serial powering chain were observed.
\\
The HDI, composed of multiple copper layers, is subject to a thermal expansion and contraction behavior that differs from that of the silicon sensor to which it is glued. Under temperature gradients, this effect induces mechanical stress especially on the bump-bond connection between the sensor and the readout-chip. To assess the bump-bonds integrity and durability, the pixel modules were exposed to thermal stress tests, up to very high temperature gradients for a selected subset of prototypes (Fig. \ref{fig:thermal_cycles}). 

% put the thermal cycles instead

\vspace{-0.15cm}
\section{Summary and Outlook}
This work summarizes the comprehensive and successful validation of the pixel module design for the CMS Phase-2 upgrade. All prototype modules and their components performed in line with expectations: the HDI layout reliably supports multiple data transmission configurations, power consumption measurements agree well with simulation results, and the Shunt-LDO powering scheme demonstrates the expected behavior. Furthermore, the pixel module design has proven to be highly robust, withstanding extensive stress tests. Prototype modules endured harsh thermal cycling without visible mechanical damage, all while maintaining stable and good performance.

\end{document}